\documentstyle[preprint,prb,aps,epsf]{revtex}
\begin{document}
\headheight2.0cm
\headsep1.2cm
\baselineskip8mm
\tolerance=1500
\renewcommand{\vec}[1]{\mathbf{#1}}
\thispagestyle{empty}
\title{Screening of the Photon Field in Surface Photoemission from
Simple Metals}
\author{S. Natschl\"ager$^1$, J. Alberich$^2$, N. Barber\'an$^2$,
J. Bausells$^3$, R. Ruppin$^4$  and C. Sim\'o$^5$}
\address{$^1$Institut f\"ur Theoretische Physik,
Johannes Kepler Universit\"at, A-4040 Linz, Austria}
\address{$^2$Departament d'Estructura i Constituents de la
Mat\`eria, Facultat de F\'{\i}sica,
Universitat de Barcelona, E-08028 Barcelona}
\address{$^3$Centro Nacional de Microelectr\'onica-CSIC,
E-08193 Bellaterra (Barcelona), Spain}
\address{$^4$SOREQ NRC, Yavne 81800, Israel}
\address{$^5$Departament de Matem\`atica Aplicada i Analisi,
Universitat de Barcelona, E-08007 Barcelona,
Spain}
\maketitle
\begin{abstract}
A previous two-step hydrodynamic model calculation
of angle-resolved
photoemission has been improved by modeling the surface with a
soft analytic function. The same ground-state potential is used to
screen the photon field and to calculate the electronic wave
functions.
The implemented code can be used for arbitrary  ground state densities
along the perpendicular direction.
Mathematical and physical aspects of the model are discussed
extensively. The behavior of the main peak of the spectrum
in response to the variation of some parameters leads us to the
conclusion that
the peak cannot be interpreted as the direct excitation of multipole
modes. This
conclusion was suggested in the previous two-step model calculation.
\end{abstract}

\hspace*{0.5cm}


\subsection*{1. Introduction}

The subject of the interaction of external probes with metal
surfaces,
and the possibility of excitation of multipole modes, have attracted
much attention. The multipole modes, which were first
predicted theoretically by Bennett \cite{ben} in 1970, from the
hydrodynamic model, have only recently been detected experimentally by
inelastic electron scattering on Na, K, Cs and Ag surfaces.
\cite{tsu1,tsu2,mor} Theoretical quantum-mechanical interpretations
of these experiments, based on the random phase approximation and on
the local density approximation, have been presented.
\cite{tsu1,tsu2,naz}
Although no multipole modes could be observed in electron scattering from Al
surfaces, it was debated whether a peak which appears in the
photoemission spectrum of Al \cite{lev,barm} could be attributed to
multipole modes. \cite{tsu2,plu,bar} To provide more insight into this question,
theoretical calculations of the angle dependence of the photoemission
spectrum are required. Here we present such a calculation, using a
refined hydrodynamic model, into which a realistic surface density
profile is incorporated.

Photoemission within a solid needs some
linear momentum supplier besides the two components:
the
incoming photon and the emitted electron.
An isolated electron cannot absorb
a photon in a process that preserves total energy and momentum.
Due to
the fact that the electrons have a finite mean free path, the
photon is absorbed
near the surface. Within this region, the most
important momentum
suppliers in a simple metal are the step potential and the screened
photon field.

These two contributions can be determined explicitly
in a Hamiltonian
formulation. The perturbation produced by the incident photon is given
by
\begin{equation}
\label{h1}
H_1=-\frac{e}{2mc}\,\,(\vec{p}\cdot\vec{A}+\vec{A}\cdot
\vec{p}),
\end{equation}
where only terms up to first order in the vector potential $\vec{A}$ are
considered and where the gauge $\Phi=0$ ($\Phi$ is the scalar
potential) has been used. $\vec{p}$ is the linear momentum
operator and $c,\,e$, and $m$ are the velocity of light in
vacuum and
the electron charge and the mass, respectively. We will employ atomic
units throughout the paper.

If $\mid i\,\rangle$ and $\mid f\,\rangle$
are two eigenstates of the unperturbed
Hamiltonian $H_0$ and $\vec{A}$ is the external plus the induced
field, the amplitude of the excitation probability
is given (apart from constant factors) by:
\begin{equation}
\label{pa}
\langle f\mid \vec{p}\cdot\vec{A}+\vec{A}\cdot\vec{p}\mid
i\,\rangle\,\,= -i
\,\langle f\mid
(\vec{\nabla}\cdot\vec{A})\,+2\vec{A}\cdot\vec{\nabla}\mid
i\,\rangle.
\end{equation}
While the first term on the right hand side is directly related
to the spatial variation
of the vector potential due to
screening effects within the electron gas, the second term is related
to the step potential at the vacuum-solid interface. This follows from
the fact that for the unperturbed Hamiltonian $H_0=p^2/2m\,+\,V$ we
have
\begin{equation}
\label{pp}
\langle f\mid \vec{\nabla}\mid
i\,\rangle\,=\,\frac{1}{E_i-E_f}\,\langle f\mid\vec{\nabla} V\mid
i\,\rangle.
\end{equation}
Therefore, if $\vec{A}$ is taken out before the integral in the
second term of the right hand side of Eq. (\ref{pa}) (as
in some approximations with optical photons), then the entire
contribution to the momentum would come from the variation of
the ground state potential at the
surface. However, in a full calculation, as $\vec{A}$ must be placed
inside the bracket, the two effects (screening and inhomogeneities of
$V$) are mixed.

These two contributions to the excitation probability are connected
with  different
kinds of processes during the electron emission: collective and
single-particle excitations respectively.

The contributions connected with the potential gradient term,
$\vec{\nabla} V $, consist of individual electron-hole pairs
created at the surface and of the subsequent emission
of the electron.
On the other hand, in the processes connected with the
divergence
of the vector potential the electron
gas
screens the incoming photon, producing surface collective excitations
that supply the extra momentum.

Moreover, different types of collective excitations play
different roles in photoemission. When the photon frequency
is equal to
$\omega_p$,
the bulk plasma frequency, the amplitude of the vector
potential is drastically reduced inside the metal and no
photoelectrons
are produced, giving a minimum in the photoemission spectrum.

Ordinary surface plasmons of frequency $\omega_p/\sqrt{2}$ cannot be
produced by an incident photon in a perfect surface, as the lightline
and surface plasmon dispersion relations
have no points in common on the
$\omega\,\,$vs.$\,\,q$ plane due to retardation effects ($q$ is the
momentum parallel to the surface).

Multipole modes associated with oscillating
charge fluctuation
in the direction perpendicular to the surface are the ideal
candidates to be the main components in the non-zero contribution
related to the divergence term.

It has been proved that, in contrast to the ordinary surface modes,
multipole modes are very sensitive to the electronic structure at the surface
\cite{ben}.

The ordinary surface modes reflect bulk
properties and at $q=0$ their energy
has only Coulomb contribution. However, even at $q=0$ multipole modes
have kinetic contribution in addition to the Coulomb part, giving
information about local properties. \cite{sel}
Several approximations of the surface region have been
published in an attempt to model the most important part of the
response of the system. Some use the hydrodynamic model
\cite{ben,kem,egu,sch}
and others use microscopic calculations \cite{fei1,fei2,sam} in
which
the photon field is screened with the same potential step as the one
used to obtain the electronic wave functions.

Here we improve a previous two-step profile calculation
\cite{bar} using a soft
electronic profile to avoid the discontinuities at the vacuum-layer
and layer-bulk boundaries that could be related with artificial
enhancement of the photoemission spectrum.

A self-consistent screening of the field is performed using an
analytical expression for the ground state density.

This paper is organized as follows. The photoemission formulation is
presented in Sec. 2. Section 3 contains the calculation of the
screened field and Sec. 4 contains the results and conclusions.

\subsection*{2. Photoemission Formulation.}

We consider a semi-infinite simple metal with its surface perpendicular
to the $z$-axis and confined to the $z<0$ half-space.
The jellium model
is assumed for the positive ionic density defined as
\begin{equation}
\label{jellium}
N(z)\,\,=\,\, \tilde{n}_0\,\,\Theta (-z)\,\,,
\end{equation}
where $\tilde{n}_0=3/4 \pi r^3_s$ is the constant bulk
density, $\Theta$ is the step function  and $r_s$ is the electron
radius.

Taking advantage of the translational symmetry on the $xy$-plane,
the photoemission flux of electrons that reach the detector per
unit of the solid angle and per incident photon can be written as
\begin{equation}
\label{dIdO}
\frac{dI}{d\Omega} \,=\,
\frac{p_{f}\,\cos(\theta)}{\omega}
\left| i(q+2p_{\| i}) \int dz\phi_f f
\phi_i\,\,+\int dz
\phi_f j \phi_i\,\,+\,\,2\int dz\phi_f g \phi^{'}_i\right|^{2}\quad,
\end{equation}
where we have followed the scheme developed by Mahan in Ref.~\onlinecite{mah}
(with the addition of the term $\vec{\nabla}\cdot \vec{A} \neq 0$, not
considered in Mahan's formulation).
In Eq. (\ref{dIdO}) $\phi_i$ is the $z$-dependent part of the initial
electronic state at the Fermi energy,
\begin{equation}
\Phi_i(\vec{r})=e^{i\vec{p}_{\| i}\cdot \vec{\rho}} \phi_i(z)\,,
\,\,\,\,\,\,\,\,\,\,\vec{r}=(\vec{\rho},z)\,\,,
\end{equation}
where $\vec{p}_{\| f}=\vec{p}_{\| i}+\vec{q}$ is the balance between
the final and initial electronic momenta parallel to the surface and
the parallel momentum $\vec{q}$ supplied by the incident photon.
\begin{equation}
\Phi_f(\vec{r})=e^{-i\vec{p}_{\| f}\cdot \vec{\rho}} \phi_f(z)
\end{equation}
comes from the Green function of a free
electron that propagates
from the metal to the vacuum through the surface
potential at a constant energy $E_f$, given by
\begin{equation}
E_f\,\,=\,\,E_i\,\,+\,\,\omega\,\,,
\end{equation}
where $E_i$ is the Fermi energy.

The functions $\Phi_i$ and $\Phi_f$ are solutions to the
Schr\"{o}dinger equation for the unperturbed Hamiltonian $H_0$.

The functions $f$ and $g$ are the $x-$ and $z-$components of the
screened electric field given by
\begin{equation}
\vec{E}=\frac{i\omega}{c}\vec{A}\,\,,
\end{equation}
and
$j=g'$ is the $z$-derivative of the $z$-component. The field
components inside the integrals ($f$,$\,\,g$ and $j$) are normalized
to
unit amplitude for the $x$-component of the incident photon in vacuum.
We consider {\em p}-polarized light ({\em s}-polarized light would not produce
a longitudinal component inside the metal) with an angle of incidence
$\theta$ with respect to the normal to the surface.

Two modifications to Eq.(2.9) in Ref.~\onlinecite{mah} have been applied to
obtain our Eq.(\ref{dIdO}). Both of them mimic the experimental
conditions used to compare our results. The first is that we
do not perform the integration over initial states as we keep the
initial state fixed and change $\phi_f$ as we scan the frequency. The
second is the normalization of the output to the incident flux
(photons/sec), given by
\begin{equation}
\frac{E_0^2\,\,\cos(\theta)}{\omega}\,\,,
\end{equation}
where $E_0$ is the amplitude of the incident photon far from the
surface in vacuum.

We stress that the electron crosses the
surface as a free particle and the full many-body effects are
considered in the dressed field.

By far the most cumbersome part of the calculation is the screening
of the photon field. We will focus on this part in the next
section.

\subsection*{3. Screening of the Photon Field.}

Our initial system of equations, within the hydrodynamic model, is
given by
\begin{eqnarray}
\label{m1}
\vec{\nabla} ( \vec{\nabla}\cdot\vec{E})-\nabla^2\vec{E}&=&-
\frac{1}{c^2}\left[\frac{\partial^2\vec{E}}{\partial t^2}+4\pi
\frac{\partial\vec j}{\partial t}\right]\quad,\\
\label{m3}
\vec{\nabla}\cdot\vec{E}&=& 4\pi(n-N)\quad,\\
\label{m2}
\frac{\partial\vec{j}}{\partial
t}&=&n\vec{E}-\beta\vec{\nabla}n-\gamma\vec{j}\quad,
\end{eqnarray}
for the three unknowns $\vec{E}$, $\vec{j}$ and $n$; $n$ and
$\vec{j}$ are the electronic density and current
respectively ($\vec{j}=n\vec{v}$), $N$ is the ionic density given by
Eq. (\ref{jellium})
, $\beta=\frac{3}{5} v^2_F$, $v_F(z)$ being the local Fermi velocity
and
$\gamma=\omega_p(z)/d$ is the damping parameter. The value $d=50$ has
been used in the present work.

In order to linearize the equations we consider
\begin{eqnarray}
\label{a1}
n&=&n_0+n_1 \exp(-i\omega t)\quad,\\
\label{a2}
\vec{E}&=&\vec{E_0}+\vec{E_1} \exp(-i\omega t)\quad,\\
\label{a3}
\vec{v}&=& \vec{v}_1 \exp(-i\omega t)\quad,\\
\label{a4}
\beta &=& \beta_0+\beta_1 \exp(-i\omega t)\quad,\\
\label{a5}
\gamma &=& \gamma_0+\gamma_1 \exp(-i\omega t)\quad,
\end{eqnarray}
which gives by substitution in the previous system,
\begin{eqnarray}
\label{e2}
\vec{\nabla}(\vec{\nabla}\cdot\vec{E_0})-\nabla^{2}\vec{E_0}&=&0\quad,\\
\label{e1}
\vec{\nabla}\cdot \vec{E_0}&=&4\pi(n_0-N)\quad,\\
\label{zero-first}
n_0\vec{E_0}&=&\beta_0\vec{\nabla}n_0
\end{eqnarray}
to zeroth order in the perturbation, and
\begin{eqnarray}
\label{e4}
\vec{\nabla}(\vec{\nabla}\cdot\vec{E_1})-\nabla^{2}\vec {E_1}&=&
\frac{\omega^{2}}{c^{2}} \vec{E_1}+\frac{4 \pi i \omega}{c^2}
n_0\vec{v_1}\quad,\\
\label{e3}
\vec{\nabla}\cdot \vec{E_1}&=& 4 \pi n_1\quad,\\
\label{e5}
-i\omega n_0\vec{v_1}&=&n_0\vec{E_1}+n_1\vec{E_0}-\beta_0 \vec{\nabla}
n_1-\beta_1\vec{\nabla}n_0
- \gamma_0 n_0\vec{v_1}
\end{eqnarray}
to first order in the perturbation.

The standard procedure would be to solve the zeroth order
system obtaining the ground state density $n_0$ from a given ionic
distribution. Instead, we proceeded in a different way: we chose a
well justified analytic electronic ground state profile and calculated
$\vec{E_0}$ from Eq. (\ref{zero-first}).

In the case of aluminum, the analytic form

\begin{equation}
\label{gsdensity}
n_0(z)=\frac{\tilde{n}_0}{e^{z/\delta}+1}\,\,,
\end{equation}
with $\delta=0.66$ a.u. agrees well with the profile obtained
from the
Kohn Sham calculation presented in Ref.~\onlinecite{ste}, see Fig. 1.
This electronic profile together with
\begin{equation}
V(z)=-\int E_0\,\,dz\,\,,
\end{equation}
where
\begin{equation}
\label{step}
\vec{E_0}=\frac{3}{5}(3\pi^2)^{2/3} \,\,n_0^{'}\,\,n_0^{-1/3}
\vec{e_z}\quad,
\end{equation}
obtained from Eq. (\ref{zero-first}),
completely characterize the ground state of the system. In Eq. (\ref{step})
$\vec{e_z}$ is a unit vector along the $z$-axis and
$n_0^{'}=dn_0/dz$.

After some manipulations in the first-order system, we obtain
the equations
\begin{eqnarray}
\label{a}
a_1f\,\,+a_2f''\,\,+a_3g'\,\,&=&0\quad,\\
\label{b}
b_1f\,\,+b_2f'\,\,+b_3g\,\,+b_4g'\,\,+b_5g''\,\,&=&0\quad;
\end{eqnarray}
the $z$-dependent coefficients $a_i$ and $b_i$ are given in the
Appendix.

This coupled system of second order differential equations
can be transformed into
an equivalent fourth order equation for one variable or, alternatively,
into a first-order system of equations for the four functions
$y_1(z)\,\,$,...,$\,y_4(z)$,
defined as $f=y_1$, $f'=y_2$, $g=y_3$ and $g'=y_4$, which satisfy
\begin{eqnarray}
\label{y1}
y'_1(z)&=&y_2(z)quad,\\
\label{y2}
y'_2(z)&=&A_1 y_1(z)+A_3 y_4(z)quad,\\
\label{y3}
y'_3(z)&=&y_4(z)quad,\\
\label{y4}
y'_4(z)&=&B_1y_1(z)+B_2y_2(z)+B_3y_3(z)+B_4y_4(z)quad,
\end{eqnarray}
which must be solved numerically. The $z-$dependent
coefficients $A_i$ and $B_i$ are defined as
\begin{equation}
\label{ab}
A_i=-a_i/a_2\,\,\,\,\,\,B_i=-b_i/b_5
\end{equation}
and are given in the Appendix. The fields were integrated from inside
using the ``odeint'' and ``stiff'' subroutines given in
Ref.~\onlinecite{pre}.
These codes change the size of the integration step according to the
nature of the profile and appeared to be the appropriate method for
our stiff profile.

The main difficulty lies in the consideration of the boundary
conditions. Deep inside the metal at $z_1<0$, where the electronic
density becomes a constant function, the four independent analytical
solutions are known:
two longitudinal and two transverse plane wave functions moving in
opposite
directions.

Let us define the coefficients $C_i$ within a general solution inside
this region as
\begin{equation}
\label{fc}
f(z)=\,C_1\,e^{-ik_1z}\,+\,C_2\,e^{-ik_2z}\,+\,C_3\,e^{ik_1z}\,+\,C_4
\,e^{ik_2z}\quad,
\end{equation}
where
\begin{eqnarray}
\label{k1}
k_1\,&=&\left[\frac{\omega^2}{c^2}\left(1-\frac{\omega_{p0}^2}{\omega(\omega+
i\gamma_0)}\right)-q^2\right]^{1/2}\quad,\\
\label{k2}
k_2\,&=&\left[\frac{\omega(\omega+i\gamma_0)-\omega_{p0}^2}{\beta_0}-q^2\
\right]^{1/2}
\end{eqnarray}
are  $z-$components of the transverse and longitudinal wave
vectors at $z_1$. The roots with Im$(k_i)>0$ have been used.

The absence of incoming waves from $-\infty$
 gives
the first two of the boundary conditions, i.e., $C_3=C_4=0$.

Far from the surface in vacuum at $z_2>0$, where the density can be
considered as a negligible constant function, the general
analytical solution is given by

\begin{equation}
\label{fd}
f(z)=\,D_1\,e^{ikz}\,+\,D_2\,e^{-ikz}\,+\,D_3\,e^{ik_3z}\, + \,D_4\,
e^{-ik_3z}
\end{equation}
where $k$ and $k_3$ are the $z-$
components of the transverse and longitudinal wave vectors
respectively. They are obtained from equations similar to Eqs.
(\ref{k1}), and
(\ref{k2}) using the
values of $\omega_p$, $\gamma$ and $\beta$ at $z_2$. We used the
condition $n(z_2)/\tilde{n_0}=10^{-8}$ and $z_1=-z_2$ to fix the $z_i$
values. The remaining two boundary conditions must be the
cancelation of both longitudinal solutions at $z_2$, or $D_3=D_4=0$.
In this way, the
allowed radiation in vacuum is the incoming and
reflected photon.

The application of boundary conditions at different points in space
does not always have a solution, as the system must be
compatible with them.
In a numerical problem, this compatibility is a problem of relative
accuracy. Taking advantage of the linear character of the
system of equations,
we inspected the $4\times4$ matrix $M$ that relates the
analytical solutions at
$z_1$ and $z_2$ and is given by
\begin{equation}
\label{d}
D=MC\quad.
\end{equation}
The result of the inspection was that
the matrix has a $2\times2$
box
of nearly zero elements,
those denoted by $m_{31},m_{32},m_{41}$ and
$m_{42}$. This box produces negligible $D_3$ and $D_4$ coefficients
for arbitrary values of $C_1$ and $C_2$.
However, we verified that the results were much more stable
if we nonetheless imposed the two last boundary conditions using
the
``shooting method''. \cite{pre}

Figures 2 and 3 show the $z-$component of the vector potential as a
function of $z$, normalized to the constant transverse component at
$z_1$. In addition to the general discussion that we will include in
the
next section, some comments about the behavior of the fields due to
screening effects in the region where the electronic density is
negligibly small must be included here.

Rapid oscillations of the $z-$component of the vector potential
appear in the region where the electronic density has nearly
disappeared (see Fig.1 for comparison). The main reason for this
unphysical behavior is attributable to the hydrodynamic model
we are using, which fails to reproduce the correct response when the
electronic density is negligible. For the longitudinal
component
(which represents the collective excitation of the system), the
dispersion relation in a homogeneous system is given by
\begin{equation}
\label{w}
\omega(\omega+i\gamma)=\omega_p^2+\beta(q^2+k_3^2).
\end{equation}

As the density goes to zero locally, $\beta$, $\gamma$ and $\omega_p$
also tend to zero. However, $\omega$ and $q$ are finite quantities
fixed by the incoming photon with which the system resonates.
The only way to satisfy this equation is by a strong
increase in $k_3$; indeed, it must grow to infinity to compensate for
the vanishing of the $\beta$ coefficient. However, the
consequences of these oscillations are not important, since the
amplitude
of the field tends to zero.  This assertion is proved in
the comparison shown in Fig. 4, as discussed in the next
Section.

This annoying effect could be avoided by the introduction of an extra
term in the dispersion relation, the well known $K^4/4$
($\vec{K}=(\vec{q},k_3)$) term, \cite{boh} related with electron-hole
excitations
(and
coming from the kinetic term within the Hamiltonian), which dominates
the dispersion relation for high energy and large momentum transfer.
If this term is included, as the ground state density tends to zero,
$k_3$ converges to a finite value given by $k_3^2=2\omega-q^2$ and
the strong oscillations would disappear.

As was recently suggested in Ref.~\onlinecite{tok} this term can be
obtained
within the hydrodynamic model for a homogeneous electron gas from
the material equation if the local equilibrium approximation is
relaxed and extra terms coming from higher momenta of the
distribution function are included. We tried with two different extra
terms within the material equation built up from the polarization
vector ($\vec{\nabla}\cdot\vec{P}=-n$), one of them free of shear
forces and given by $\vec{\nabla}(\nabla^2 n)$, and the other one
containing shear forces and given by $(\nabla^2)^2 \vec{P}$.
These two terms include all the possible vector components that
contain four derivatives of $\vec{P}$. They both imply the
dispersion relation given by:
\begin{equation}
\label{dispnew}
\omega(\omega+i\gamma)=\omega_p^2+\beta(q^2+k_3^2) +\frac{1}{4} (
q^2+k_3^2)^2
\end{equation}
which reproduces the random phase approximation result of
Ref.~\onlinecite{hed}. Both
produce a new system of six coupled first-order equations which
need six boundary conditions. The difficulty comes from the
fact that
two of the four solutions of $k_3$ in Eq. (\ref{dispnew}) are
unphysical, as for them
Im($k_3$)$\cdot$Re($k_3$)$<0$, which corresponds to a plane wave
that diverges in the forward direction producing divergent solutions
as the differential equation is solved from $z_1$ to $z_2$.
The goal of our future work is to find the appropriate term in the
material equation
that would leave the system linear and simultaneously
produce well-behaved solutions of the
dispersion relation.

\subsection*{ 4. Results and Conclusions.}

The physical parameters of our problem are the width of the surface
profile
given by $\delta$, the angle $\theta$ between the incident photon and
the normal to the surface, the angle $\epsilon$ between the outcoming
electron and the normal to the surface and the electron radius $r_s$.

In Figs.2 and 3 the $z-$component of the vector potential is shown as
a function of
$z$
for different values of $\omega$.
 It has been normalized to
the transverse component at $z_1$ and consequently the deviation from
$1$ (of the real part) gives an idea of the importance of the
longitudinal
component deep inside the metal. It can be observed that even though
Im($k_2$)$\gg$ Im($k_1$) and so strong damping is expected,
its presence is significant at large distances from
the surface and cannot be ignored for $\omega>\omega_p$.

Some characteristic features that were previously reported by
Feibelman \cite{fei2} within a self consistent
calculation are reproduced here by us:
$\omega=\omega_p$ is the critical
value from which the wave vector of the longitudinal component has
a nonvanishing
real part and consequently it penetrates the
metal, producing decaying oscillations towards $-\infty$. For this
value of
$\omega$ the amplitude of the nearly structureless field shows an
abrupt decrease, producing a
minimum (it should be zero if $\gamma=0$, i.e. if no damping is
considered) in the photoemission spectrum.
For $\omega<\omega_p$ there is a strong peak near $z=0$ produced by
screening, a feature that strongly deviates from the classical
optical calculation.
Correlating the strength of the surface photoeffect with the
intensity of the vector potential within the surface region, one
would expect it to be large below $\omega_p$ and much smaller
above $\omega_p$.

Indistinguishable results were obtained when the
selfconsistent density shown in Fig.1 was used.

In order to see the influence of the oscillations of the field at
the vanishing density region on our results, we compare in Fig.4
for $\delta=0.2$ a.u. (which mimics an abrupt electronic profile)
the
evolution with $\omega$ of the two transmission coefficients, $C_1$
(transverse component)
and $C_2$ (longitudinal component) and the reflection coefficient
$D_1$ with those obtained
analytically from a  calculation that assumes a single step,
structureless
surface. The analytical coefficients are given in
Ref.~\onlinecite{mel} by
\begin{eqnarray}
\label{c1}
C_{1c}&=&\frac{2k}{t}\frac{k_1}{q}\quad,\\
\label{c2}
C_{2c}&=&\frac{2k(\epsilon
-1)}{t}\frac{q}{k_2}\quad,\\
\label{c3}
D_{1c}&=&\frac{\epsilon
k-k_1+(\epsilon-1)q^2/k_2}{qt}\sqrt{\frac{\omega
^2}{c^2}-q^2}\quad,
\end{eqnarray}
where
\begin{eqnarray}
\label{c4}
t&=&\epsilon k+k_1-(\epsilon-1)q^2/k_2\quad,\\
\label{c5}
\epsilon&=&1-\frac{\omega^2_p}{\omega(\omega+i\gamma)}\quad,
\end{eqnarray}
and $k,k_1,k_2$ are defined in Eqs. (\ref{k1}), (\ref{k2}) and
(\ref{fd}).
The curves do not show any
structure, except at
$\omega_p$
and at $\omega_p/\cos(\theta)$, where the transverse component can
penetrate the solid.
The coincidence of the two results proves that
the strong oscillations at the edge region
shown in Figs. 2 and 3 do not affect the behavior of the field
amplitudes as functions of $\omega$.

The same coincidence is obtained if $r_s=6$ is used, keeping
all the other parameters fixed.

Figure 5 shows the change of the
longitudinal transmission coefficient
$C_2$ [see Eq. (\ref{fc}) ] as the electronic profile becomes less
abrupt. As
$\delta$ increases,
giving a flatter profile, a wide peak at about $\omega=0.75$
$\omega_p$
evolves. Apparently, the position of the peak is nearly insensitive to
the changes in the electronic profile.

Finally, we centered on the study of the photoemission intensity per
unit of the solid angle and incident photon for {\em p}-polarized light as a
function of the photon frequency. The initial
state $\Phi_i$ is at $E_i=-0.1615$ a.u. . This energy was chosen to
reproduce the threshold value of $\omega$ for aluminum. We take the
zero of energy to be at vacuum level.
No final state effects are taken into account. This last
approximation
is supported by experiment as no qualitative
difference
is observed between photoemission spectra made on different surfaces
of aluminum crystals, suggesting that in a nearly free electron metal
the screening effects are much greater than the structure
produced by band effects. \cite{lev}

The finite mean free path of the emitted electron inside the metal
has been modeled by a decreasing exponential factor multiplying the
initial electronic wave function which ensures that only electrons
coming from the last $50$ \AA $\,\,$ will be considered.

Figure 6 shows the main result that must be compared with the
experimental outputs, \cite{lev} (for $\theta=\pi/4,\,\,\epsilon=0$
and
$\delta=0.66$ a.u.).
The main peak
is centered at $\omega=0.77\,\,\omega_p$, in excellent agreement with
the experimental peak position. The second peak, at $\omega>\omega_p$,
has a shape comparable to the experimental one, though it has
greater intensity. This too high intensity is caused by the
absence of
electron-hole excitation processes from the hydrodynamic model. Thus,
our model lacks the Landau damping mechanism, through which the
collective excitations can decay into electron-hole pairs. This type
of damping would have a significant effect only for $\omega\,>\,\omega
_p$, where the bulk plasmons created by the incident photon can overlap
the electron-hole band and disappear, thus causing the photoemission
efficiency in this region to decrease. Unlike the surface modes at
$\omega\,<\,\omega_p$, these bulk plasmons can have a high enough
perpendicular momentum, to enable this process.

More recent experimental data reported in Ref.~\onlinecite{barm} show the
main peak at $\omega=0.85$ $\omega_p$.

To gain insight into the physical significance of the main peak,
we fixed $\delta=1.2$ a.u., $\epsilon=0$ and $r_s=2.07$ and looked
for the variation of the spectra for different values of the direction
of the incident photon (given by $\theta$). A large value of $\delta$
was chosen to increase the probability of mulitpole modes. As shown
in Fig. 7, the peak position turns out to be sensitive to the value of
$\theta$. The peak disperses to lower energies as
$q_{\parallel}$ increases.

A peak produced by a multipole mode should hardly disperse
(for these small values of $q$) and, in any case, according to previous
calculations, \cite{tsu2} it would disperse in the opposite direction.
Our results indicate that the main peak need not be a
direct manifestation of multipole modes, as has been asserted
in the literature, \cite{plu} using the coincidence in energy as the
main argument.

Another known difficulty concerning the identification of the
main photoemission peak with the multipole modes,
is that the RPA calculation of
the photoemission spectrum
\cite{fei2}
predicts that the frequency of this peak should increase with $r_s$.
However, the results of measurements of the multipole modes
frequency by electron
loss spectroscopy do not follow this trend.

In order to understand the origin of the $\theta$ dependence of
$dI/d\Omega$, let us analyze the different contributions to Eq.
(\ref{dIdO}). On the right hand side, the first two terms come from
$\vec{\nabla}\cdot \vec{A}$ and the third comes from $\vec{A}\cdot
\vec{\nabla}$. If $\vec{A}$ is a constant field, only the third term
will be nonzero and the $\theta$-dependent term will be
a global factor that will change the intensity of $dI/d\Omega$ for
different $\theta$ values, but not its shape. If $\vec{A}$ is not a
constant field, the three terms will survive and as their relative weights
are $\theta$ dependent, changes on  $\theta$ also produce changes
in the shape of the $(dI/d\Omega)/\omega$ function.

Moreover, though the contribution of the first two terms is
significant, the greatest contribution comes from the third term,
as shown in Fig. 6 (dashed line), where the first two terms of
Eq. (\ref{dIdO}) have been removed.

We now compare our results with those of two sets of experimental
data, in which the angular dependence of the photoemission was
displayed. In the case of indium, \cite{jez} our calculated results
for the main peak agree reasonably well with the measured ones. The
measured peak moves from about $0.84$ $\omega_p$ to about $0.92$
$\omega_p$, as $\theta$ changes from $45^{\circ}$ to $15^{\circ}$,
whereas the corresponding calculated peak position moves from $0.77$
$\omega_p$ to $0.92$ $\omega_p$ for the same angle variation. These
measurements were not extended to $\omega\,>\,\omega_p$, so we cannot
perform any comparisons in this region. For Al, on the other hand, the
recent experimental data of Barman \textit{et al.} \cite{barm} cover the
frequency regions both above and below $\omega_p$. Experimentally, the
main peak, bellow $\omega_p$, exhibits no dispersion when $\theta$
changes, unlike the calculated results. For $\omega\,>\,\omega_p$
there appear in the measured data two minor peaks (called features A
and B in Ref.~\onlinecite{barm}), which also occur in our calculated results.
The arrows in Figs. 6 and 7 denote the positions of the high energy
peaks, which disperse in the same way as the experimental data (feature
A of Ref.~\onlinecite{barm}). The vertical line at about $\omega/\omega_p\,=\,
1.07$ denotes the position of the minor peak (for $\theta
\,=\,50^{\circ}$ and $\theta\,=\,70^{\circ}$) which does not disperse
(feature B in Ref.~\onlinecite{barm}).

In conclusion, we have presented an analytical calculation of the
photoemission spectrum and its dependence on the photon angle of
incidence. The screening of the photon field, including nonlocal
effects and excitation of longitudinal modes, was calculated within a
hydrodynamic model with a realistic surface density profile. Our
results indicate that the main peak, below $\omega_p$, need not
be just due to mulitopole mode excitation. The two sets of
experimental data with which we compared our results seem to be
contradictory as regards the main peak dispersion. Thus, further
experiments on other materials are needed to resolve this question.

\subsection*{Appendix A}
\label{Appendix A}

Here we give the definitions of the coefficients $a_i$ and $b_i$:
\begin{eqnarray}
  \label{eq:coeffs}
  a_1(z) &=& \omega^2(\omega + i\gamma_0(z)) - \omega[4\pi n_0(z) +
  q^2\beta_0(z)]quad,\\
  a_2(z) &=& c^2(\omega + i\gamma_0(z))quad,\\
  a_3(z) &=& i q [\omega\beta_0(z) - c^2(\omega + i\gamma_0(z))]quad,\\
  b_1(z) &=& i\omega q
  \left[\frac{2}{3}\beta_0(z)\frac{n_0'(z)}{n_0(z)}-E_0(z)\right]quad,\\
  b_2(z) &=& a_3(z) quad,\\
  b_3(z) &=& -4\pi n_0(z)\omega - (\omega + i\gamma_0(z))(c^2q^2-\omega^2)quad,\\
  b_4(z) &=& \omega \left[\frac{2}{3}\beta_0(z)\frac{n_0'(z)}{n_0(z)}-
    E_0(z)\right]quad,\\
  b_5(z) &=& \beta_0(z)\omega\quad,
\end{eqnarray}
\begin{eqnarray}
\gamma_0&=&\frac{(4\pi n_0)^{1/2}}{50}\quad,\\
\beta_0&=&\frac{3}{5}(3\pi ^2n_0)^{2/3}\quad.
\end{eqnarray}

The coefficients $A_i$ and $B_i$ defined in Eq. (\ref{ab}) are given
by
\begin{eqnarray}
  \label{eq:coeffsratios}
  A_1(z) &=&  -\frac{\omega^2}{c^2}\left[ 1 -
    \frac{4\pi n_0(z)+q^2\beta_0(z)}{\omega(\omega+i\gamma_0(z))}\right]\quad,\\
  A_3(z) &=& -i q\left[\frac{\omega\beta_0(z)}
    {c^2(\omega+i\gamma_0(z))}-1\right]\quad,\\
  B_1(z) &=& -i q\frac{1}{3\delta\tilde n_0}
  n_0(z)e^{z/\delta}\quad,\\
  B_2(z) &=& -i q\left[1-\frac{c^2}{\omega}\,\,
    \frac{\omega+i\gamma_0(z)}{\beta_0(z)} \right]\quad,\\
  B_3(z) &=& \frac{20\pi}{3(3\pi^2)^{2/3}} n_0(z)^{1/3}+
  \frac{\omega(\omega + i\gamma_0(z))}{\beta_0(z)}
  \left(\frac{c^2q^2}{\omega^2}-1\right)\quad,\\
  B_4(z) &=& -\frac{1}{3\delta\tilde n_0}
  n_0(z)e^{z/\delta}\quad.
\end{eqnarray}

\subsection*{ Acknowledgments}

This work was supported in part by the Comisi\'on Asesora de
Investigaci\'on Cient\'\i fica y T\'ecnica (CAICyT), Spain under Grant
No. PB95-1249, DGICYT PB 94-0215 and CIRIT 1998SGR-00042. One of
the authors (S.N.) wishes to express his gratitude for the warm
hospitality during his stay at the Department ECM, Fisica, Universitat
de Barcelona.

\vskip3mm

\eject
\begin{list}{}{\setlength{\leftmargin}{7mm}\labelwidth1cm
\baselineskip8mm}
\item FIGURE CAPTIONS
\item[Fig. 1] Ground state density for aluminum. The solid line is the
self consistent calculation taken from Ref. \cite{ste} and the dashed
line is the analytical function given in Eq. (\ref{gsdensity}) for
$\delta=0.66$ $ a.u.$. All figures are calculated for $r_s=2.07$
$a.u.$.
\item[Fig. 2] (a) Real and (b) imaginary parts of the
$z$-component
of the vector potential $vs.$ $z$ calculated for different values of
$\omega\leq\omega_p$. The normalization constant
$A^{T}_{z}$ is the transverse component at $z_1$. $\delta=0.66$
$a.u.$, and $\theta=\pi/4$ were used.
\item[Fig. 3] Same as Fig. 2 for $\omega\geq\omega_p$.
\item[Fig. 4] Variation of the real and imaginary parts of the
coefficients $C_1$,
$C_2$ and
$D_1$
(see Eqs. (\ref{fc}) and (\ref{fd})) with $\omega$ for
$\delta=0.2$ $a.u.$, and $\theta=\pi/4$.
Insets: the same coefficients for a structureless surface.
\item[Fig. 5] Variation of the real part of the longitudinal
coefficient $C_2$ with
$\omega/\omega_p$ for different values of $\delta$. $\theta=\pi/4$
was used.
\item[Fig. 6] Photoemission spectrum as a function of
$\omega/\omega_p$ for $\delta=0.66$ $a.u.$, $\epsilon=0$ and
$\theta=\pi/4$. A partial contribution to the spectrum is included as
a dashed curve (see text). The arrow is at $\omega_p/\cos(\theta)$.
\item[Fig. 7] Variation of the photoemission spectrum with
$\theta$ for $\delta=1.2$ $a.u.$ and $\epsilon=0$. The arrows are at
$\omega_p/\cos(\theta)$.

\end{list}
\vfill
\eject

\newpage
\begin{figure}
\epsfxsize=6.0truein
{\centerline{\epsffile{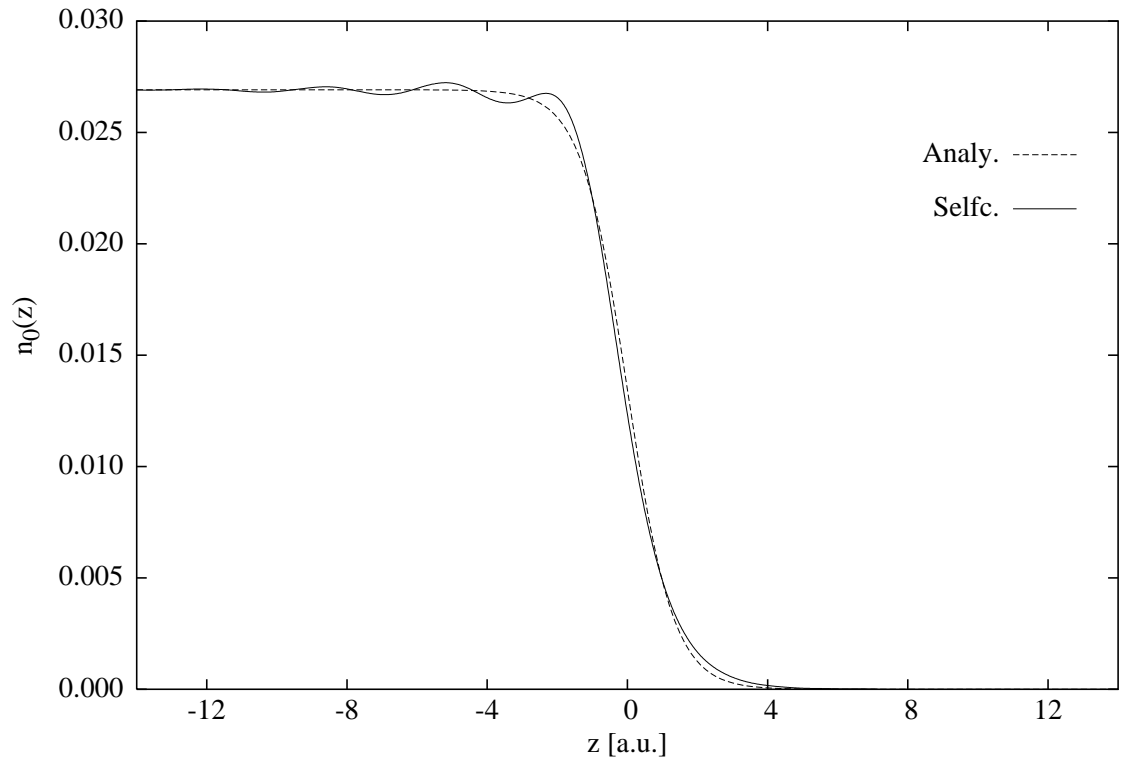}}}
\vspace{2truecm}
\caption{Ground state density for aluminum. The solid line is the
self consistent calculation taken from Ref.\protect\onlinecite{ste} and the dashed
line is the analytical function given in Eq. (\ref{gsdensity}) for
$\delta=0.66$ $ a.u.$. All figures are calculated for $r_s=2.07$
$a.u.$. 
}
\end{figure}

\newpage
\begin{figure}
\epsfxsize=5.0truein
{\centerline{\epsffile{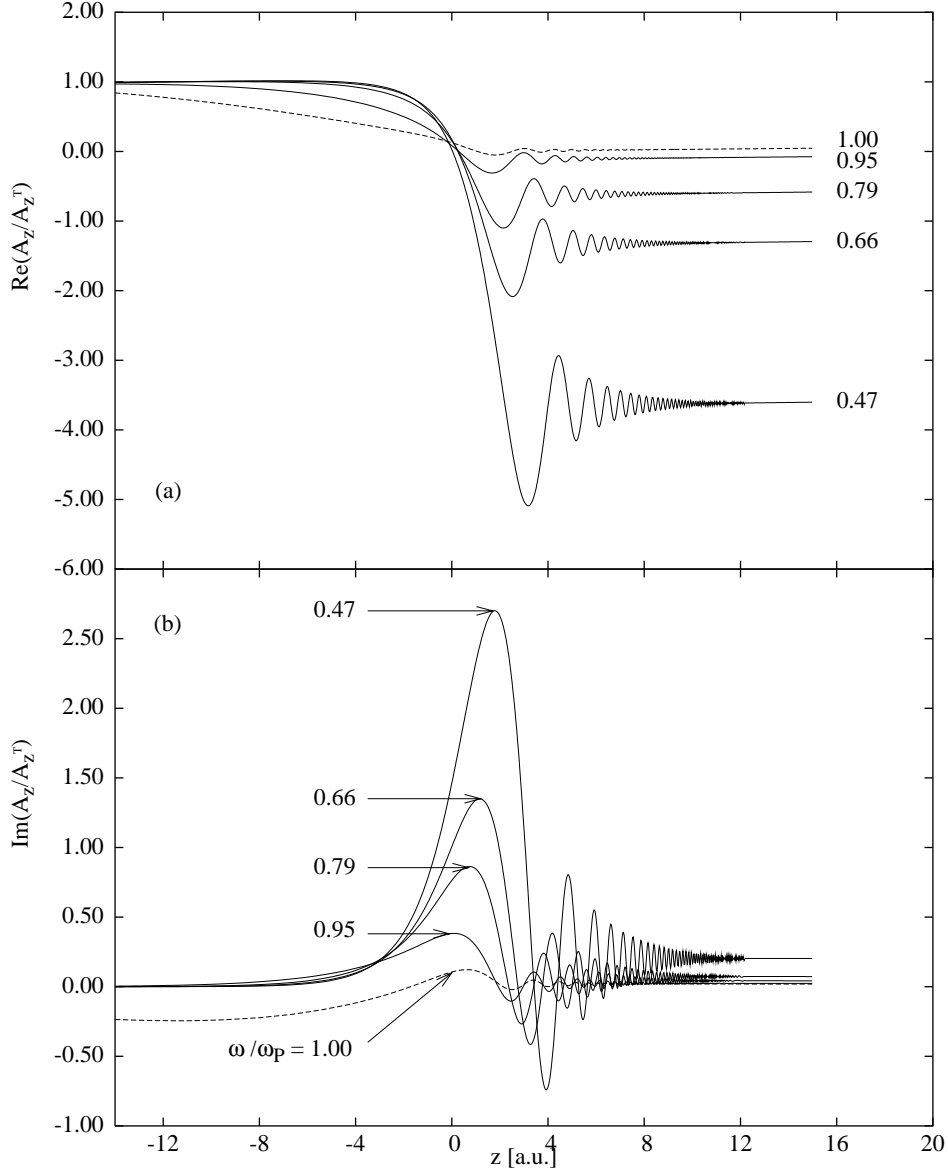}}}
\vspace{1truecm}
\caption{(a) Real and (b) imaginary parts of the
$z$-component
of the vector potential $vs.$ $z$ calculated for different values of
$\omega\leq\omega_p$. The normalization constant
$A^{T}_{z}$ is the transverse component at $z_1$. $\delta=0.66$
$a.u.$, and $\theta=\pi/4$ were used.}
\end{figure}

\newpage
\begin{figure}
\epsfxsize=5.0truein
{\centerline{\epsffile{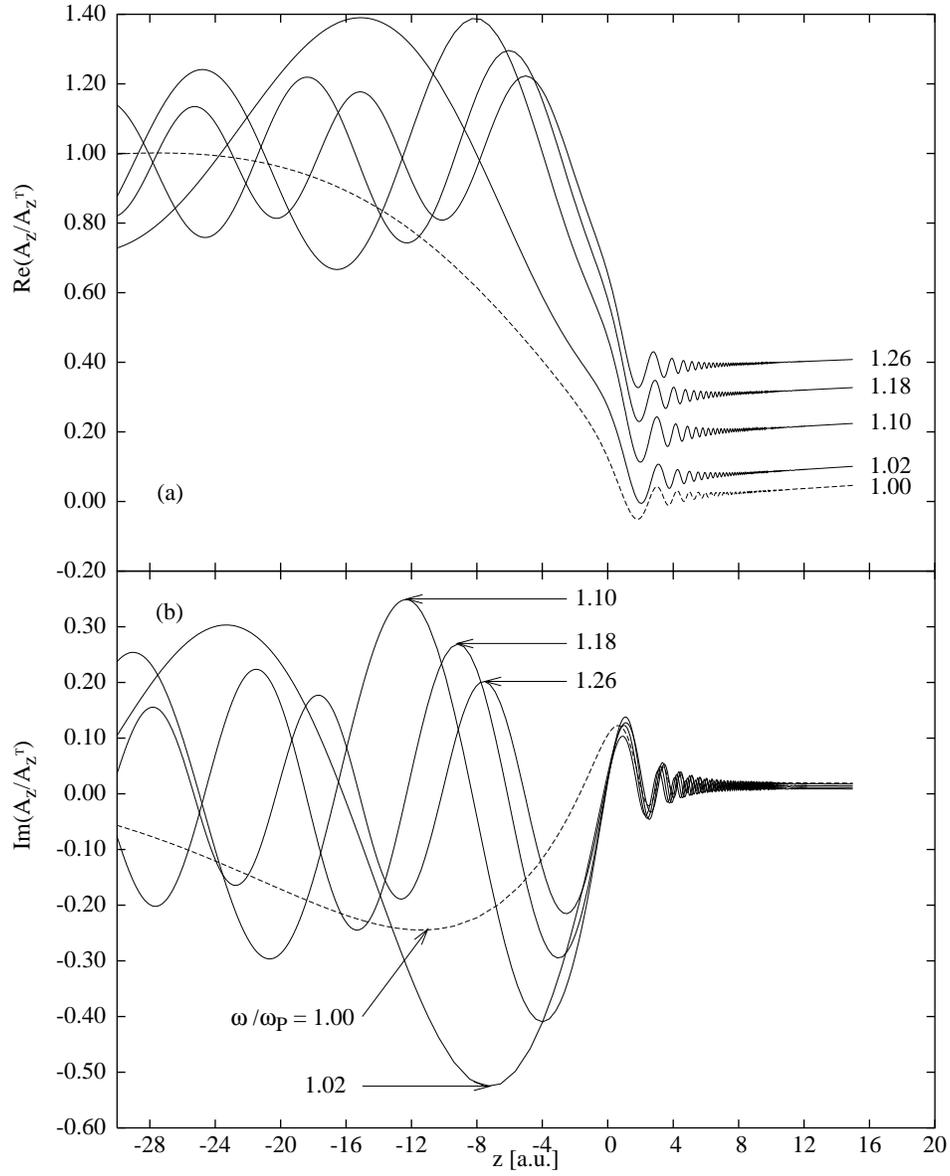}}}
\vspace{1truecm}
\caption{Same as Fig. 2 for $\omega\geq\omega_p$.}
\end{figure}

\newpage
\begin{figure}
\epsfxsize=5.0truein
{\centerline{\epsffile{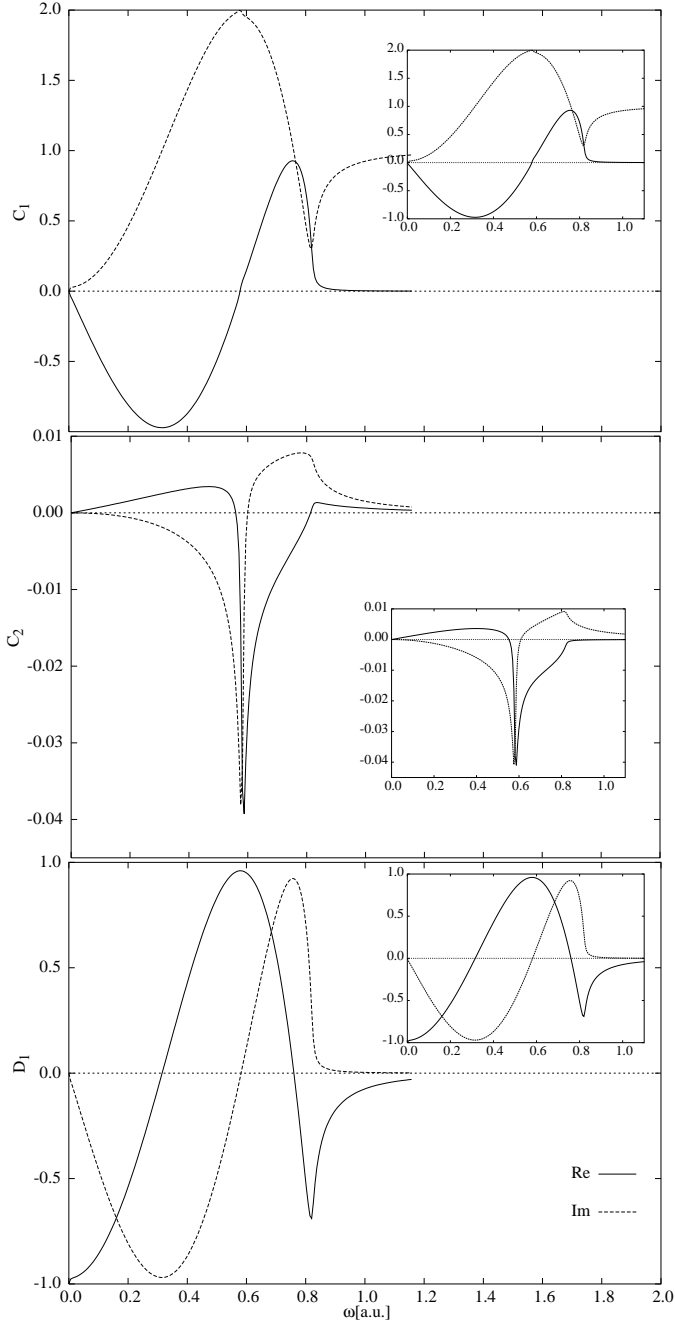}}}
\vspace{1truecm}
\caption{Variation of the real and imaginary parts of the
coefficients $C_1$,
$C_2$ and
$D_1$
(see Eqs. (\ref{fc}) and (\ref{fd})) with $\omega$ for
$\delta=0.2$ $a.u.$, and $\theta=\pi/4$.
Insets: the same coefficients for a structureless surface.}
\end{figure}

\newpage
\begin{figure}
\epsfxsize=5.0truein
{\centerline{\epsffile{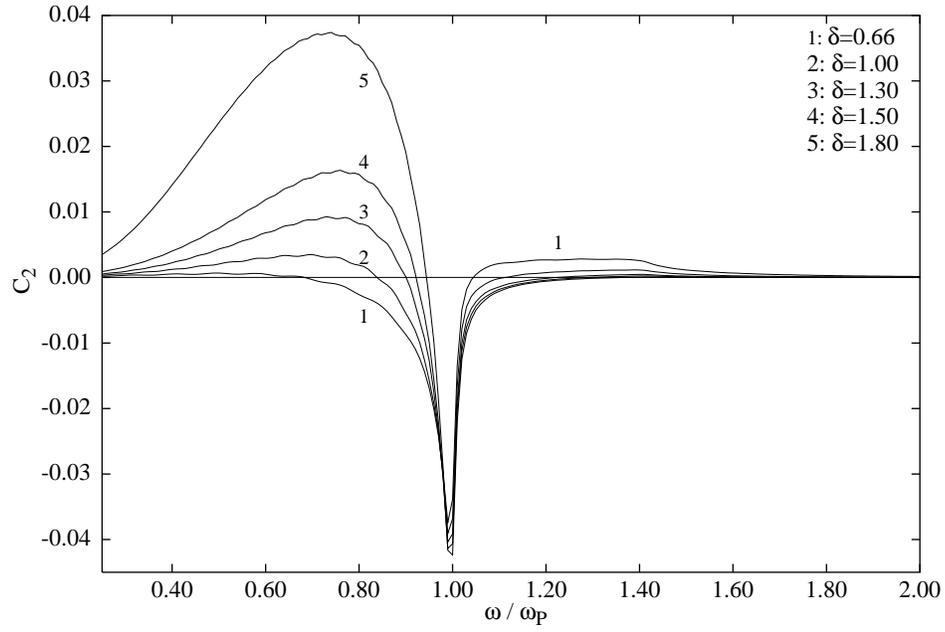}}}
\vspace{1truecm}
\caption{Variation of the real part of the longitudinal
coefficient $C_2$ with
$\omega/\omega_p$ for different values of $\delta$. $\theta=\pi/4$
was used.}
\end{figure}

\newpage
\begin{figure}
\epsfxsize=5.0truein
{\centerline{\epsffile{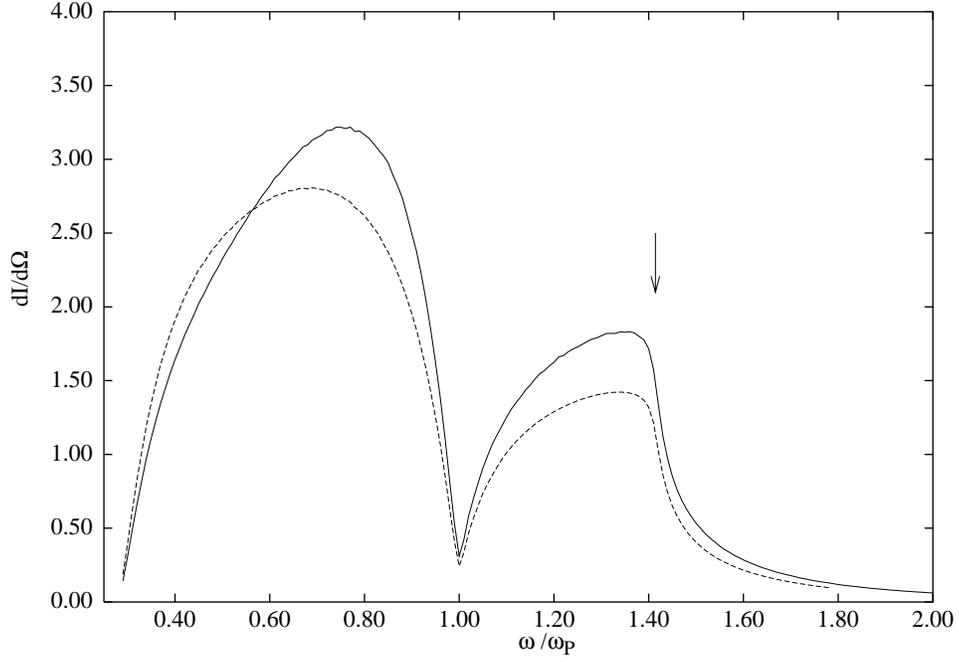}}}
\vspace{1truecm}
\caption{Photoemission spectrum as a function of
$\omega/\omega_p$ for $\delta=0.66$ $a.u.$, $\epsilon=0$ and
$\theta=\pi/4$. A partial contribution to the spectrum is included as
a dashed curve (see text). The arrow is at $\omega_p/\cos(\theta)$.
}
\end{figure}

\newpage
\begin{figure}
\epsfxsize=5.0truein
{\centerline{\epsffile{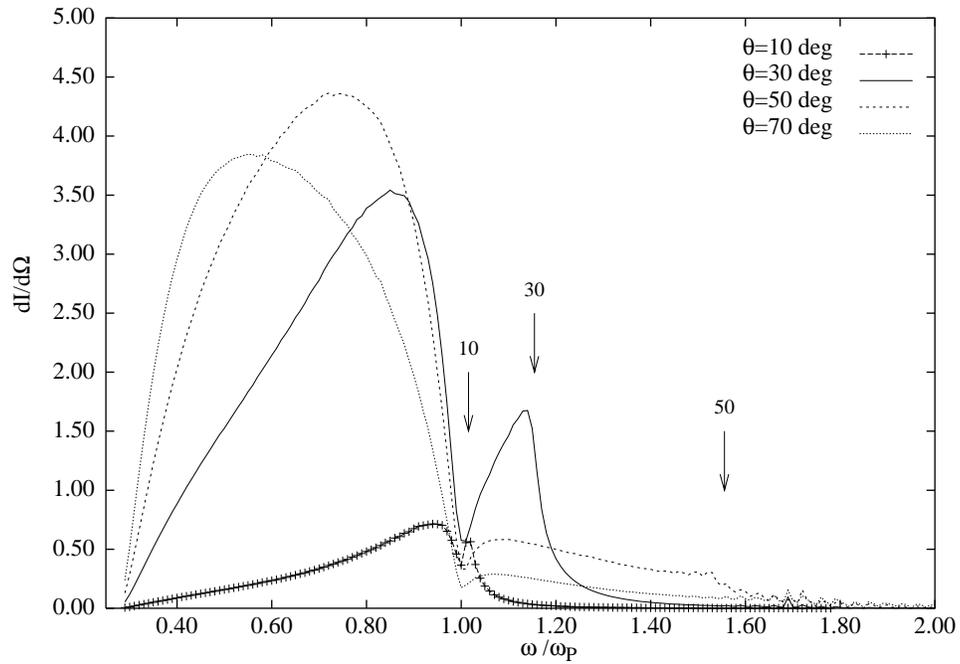}}}
\vspace{1truecm}
\caption{Variation of the photoemission spectrum with
$\theta$ for $\delta=1.2$ $a.u.$ and $\epsilon=0$. The arrows are at
$\omega_p/\cos(\theta )$.
}
\end{figure}

\end{document}